# Grid Computing Model for Mobile: A Better Mobile Grid Computing Model.

Robail Yasrab
School of Computer Scienvce.
University of Science and Technology of China

**Abstract**

Grid Computing is an idea of new kind of network technology in which research work in progress. There is a great deal of hype in this technology based area for that reason; it is getting a great deal of attention of computing community. Grid Computing makes use of clusters of PCs (personal computers), network servers or other computing technology based systems. These systems are connecting jointly to deal with complex calculations, communication and collaboration. Grid Computing offers a great deal of support for offering corporations to make use of their idle computing power, or system's processing cycles, to produce a type of supercomputer arrangement.

The technology of grid computing has newly transferred from customary high performance plus distributed technology based computing systems to utility and pervasive based computing arrangements. These technology based structures are having superior potentials of the wireless networks communication and collaboration as well as having thin, lightweight and remotely accessible devices. This has as produced the establishment new computing technology paradigm known as Mobile Grid Computing.

(Litke, Skoutas, & Varvarigou, 2004)

In grid computing paradigm, mobile or wireless based grids computing expand the potential of grid computing by means of wireless technology based devices. Presently, all through world usage of the PDAs, laptops, cell phones, and other wireless computing devices is growing that is leading to more networked wireless systems, as well as producing a gigantic collective power of unexploited resources. Mobile grid computing by means of its model of synchronized resource sharing can offer means to make use of similar resources that are general distributed all through a computing grid. We can expect to have a Grid-Net in coming future as we presently have Internet technology nowadays.

This research is going to assess and analyze the grid computing and mobile computing technology based arrangements. Here I will assess the feasibility analysis of the mobile grid computing. This research will analyze the mapping of grid computing techniques on mobile domain. In this scenario, I will assess how can, we achieve grid architecture in mobile devices. For sake of implementing a mobile gird arrangement I will analyze the existing grid computing models and discuss the design, implementation, architecture and its limitations. This research will also highlight the effect of grid computing on mobile's load balancing, high processing efficiency and less network communication.

## Chapter 1: Introduction

1) **Introduction**

*1.1- Background*

Computing Technology has extensively developed and evolved for lots of years now were governed through the law of exponentials. Extensive growth has arisen in computer storage, microprocessor speeds as well as optical capability. Technology that was previously working as an independent technology now comes together to establish a dominating technology like grid computing technology that is extremely flourishing these days (Abbas, 2003). Nowadays, we are able to carry more information and data traffic per second, on a single thread of fiber, as compared to the entire of the traffic on the entire Internet in a month in year 1997. According to Abbas (2003) this technology based developed is able to be established through the establishment of the fiber optics as well as associated technologies. These new technology based progresses allows us to place extensive data onto optical fiber (Abbas, 2003)**.**

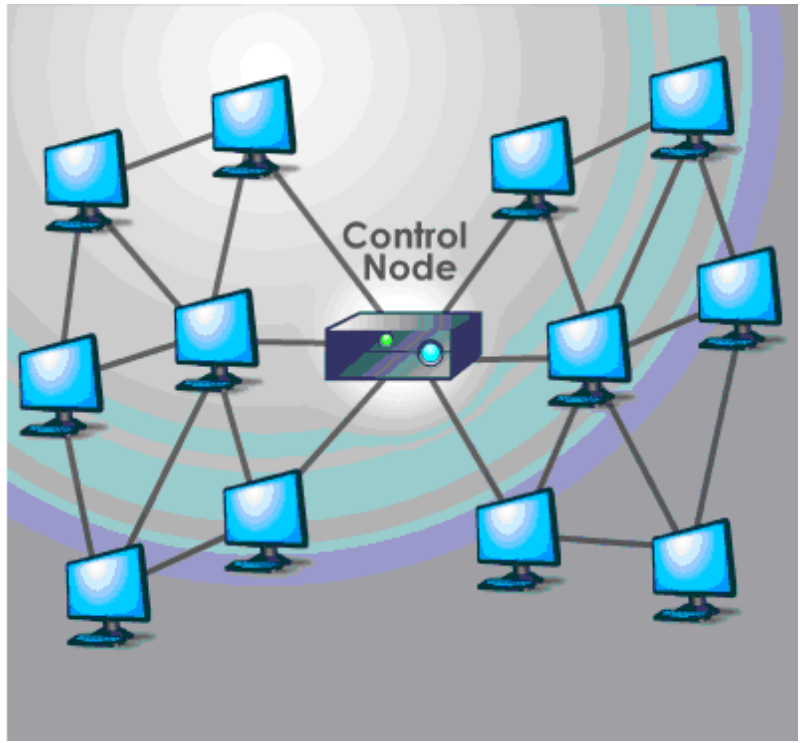

**Figure 1 Basic grid computing system**

Source: http://computer.howstuffworks.com/grid-computing.htm

### *1.2- Grid Technology Evolution*

Due to extensive development and invention in grid computing and wireless technology based systems the mobile grid computing is developing every day toward more and more enhanced technology based arrangement. For this reason speedy expansions and developments in grid computing and wireless technology are presently available. Mobile grids technology enhances the potentials of grid computing to extensively available wireless devices. Birje, Manvi, & Bhanuprasad (2006) stated that mobile grid computing are offering enhanced capability of data and information sharing by means of fixed or mobile wireless devices inside the virtual organizations. It can comprise devices similar to PDAs, laptops, sensors, mobiles etc., as the resources of these technology based devices can be memory, processor, code repositories, bandwidth, applications, etc  (Birje, Manvi, & Bhanuprasad, 2006).

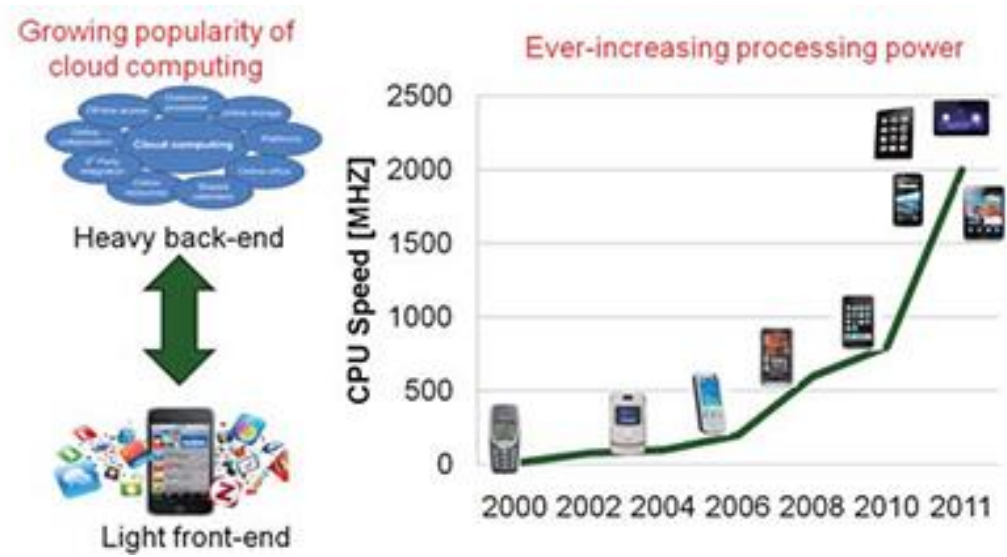

**Figure 2 Increase in Speed of Mobile Devices**

Source: http://www.isgtw.org/sites/default/files/img_2011/CACsummer2011pic1.png

Normally wireless systems are distinguished through minimized secondary storage, kless CPU performance, sensitive battery utilization, as well as unpredictable low-bandwidth communication. However these systems as well can comprise a number of extra devices as microphones, cameras, RFID readers, bar code, satellite receiver, GPS receivers and large range of unique purpose sensors. Nowadays there are variety sensors those are offering great deal of support for assessing air pressure, temperature as well as humidity. Some wireless systems are having support for senescing movement, and those that calculate radiation as well as particulate levels. Generally characteristics of wireless grid are able to be articulated as follows (Birje, Manvi, & Bhanuprasad, 2006):

No centralized control.

- Unstable and Dynamic resources and users interaction.

- Comprises various applications, resources as well as interfaces.

- New kinds of resources similar to GPS trackers, cameras plus sensors, those are able to be shared amongst grid devices and users.

- Composed of low and small powered computing devices.

- Systems are geographically detached and resources management is done through predefined policies.

- Dissimilar security needs as well as policies applications.

The cause intended for bigger popularity, scope and practice of mobile wireless systems is their small cost and size, simple management, and biggest factor is mobility. The system's mobility helps to user for the sake of visibly sharing the mobile computing grid resources, perform their jobs, as well as get instantaneous outcomes as they are roaming. Fast developments in reduction of size, growing system's processing power as well as incorporating more and more feature, in conjunction with the explosion of wireless access points have rapidly extended the value of these devices as well as formulated them more and more competent of participating in grid computing networks. As Becker, (2005) stated that the fast reduction in rates of wireless devices, attached by the simplicity of installation intended for fixed position utilize, has formulated them a gainful choice for small company and the wide-ranging public intended for mobile utilize plus for home networking (Becker, 2005).

According to Ahuja & Myers (2006) wireless mobile communication devices have turn out to be a vital tool intended for corporations of small and large size, particularly those where workers have to carry out their jobs away from the office like that sales representatives and field staff (Ahuja & Myers, 2006). The growing dependence on these systems has augmented the speed at those applications for these systems are developed, and increasing the functionality and scope of these systems. As further of these systems are installed, more power employs intended for them are being acknowledged. As the Grid is the most excellent method to attain large-scale resource distribution in a diverse situation, the mobile communication community has to experience a lot of issues to enter this systems domain. The combination of mobile communication systems is tremendously challenging for the reason that of the lot of defects that mobile devices and wireless communication have. Wireless communication networks are of low reliability and bandwidth, as mobile devices have a great deal fewer technology resources accessible as compared to desktop PCs. As Manvi (2010) stated that function of the Grid is to offer a greatly accessible, dependable as well as ubiquitous atmosphere, incorporate mobile system into the Grid environment, by means of wireless connections, could mortify the entire of the applicable quality of service needs (Manvi, 2010).

### *1.3- Mobile Grid Computing*

One of the biggest and impotent aspect for considering and assessing Mobile Grid computing is to have a reliable and perfect description, or at any rate analysis of what a

Mobile Grid is. There are numerous efforts intended for the exact description of the Grid Computing architecture. Yet a variety of methods are formulated to tackle an enhanced degree of precision the term Grid. According to Gridforum (2012) the Grid Computing arrangement is able to be observed as a high performance, distributed and collaborative data handling structure, that includes organizationally and geographically isolated, varied resources (storage systems, computing systems, tools as well as other real time human associates, data sources, contact systems) plus offers regular communication arrangement for entire of these resources, by means of open, standard, general purpose interfaces and protocols (Gridforum, 2012). Though, it is as well the foundations and facilitates technology for pervasive plus utility computing because of capability of being extremely heterogeneous, open and scalable.

Mobile Computing base arrangement is a general idea that is explaining the functions of portable, small as well as wireless communication and computing devices. This comprises devices similar to Laptops by means of WLAN (or Wireless Local Area Networks) communication technology, PDAs and mobile by means of Infrared Data Association or Bluetooth data transfer based interfaces. The Mobile Computing based arrangement spotlights on the need of offering access to communications, information as well as services all over the place, anytime as well as through some accessible means (Unger & Haynos, 2003). The technical based systems for accomplishing this are not for all the time easy to put into practice.

Actually, mobile computing necessitates the formation of communication structures as well as the change of computer OS, networks and applications. According to Litke A. , Dimitrios, & Theodora (2004) the mobility concerns involve a number of restrictions that have to be tackled since they bound the potential of a moving resource in difference to a fixed one. Mobile Grid offers a great deal of significance to both Mobile and Grid Computing arrangements. This technology based paradigm is involving a complete inheritor of Grid by means of the extra characteristic of facilitating mobile clients as well as resources in a faultless, visible, protected plus well-organized way. It has the capability to set-up fundamental ad-hoc networks as well as offers a self-configuring Grid arrangement of mobile resources (users and hosts) linked through wireless connections and establishing random and arbitrary topologies (Litke A. , Dimitrios, & Theodora, 2004).

### 1.4- Research Motivation

In this section I am going to assess the basic research motivation and application as well as significance of mobile grid technology arrangement. Grid is previously being effectively utilized in numerous scientific implementations where enormous amounts of information have to be processed and stored. In case of such challenging applications arrangement, the idea of Grid is developed, justified and diffused among the scientific communities all through the world. Bruneo, Scarpa, Zaia, & Puliafito (2003) as the amount of possible Grid clients is actually huge, the gathered data handling, processing as well as storage needs are at any rate comparable. Wireless devices (for example PDAs, mobile or laptops), by means of at present restricted resources (like low battery power, low processing power, storage constraints), would profit from the advantages of making use of significant amount of resources formulated accessible through the entire of other devices linked through network (Bruneo, Scarpa, Zaia, & Puliafito, 2003). As Gai (2012) stated especially, mobile users might be the potential clients of this new technology of GRID systems. Furthermore, we have traveling users who travel as well as work just seldom at their workplaces (Gai, 2012).

As Park, Ko, & Kim (2003) stated Mobile Grid facilitates equally the resources inside the GIRD computing arrangement that are themselves component of the Grid as well as mobility of the clients requesting access to a fixed amount of resources. These both situations have their own restrictions and limitations that should be managed (Park, Ko, & Kim, 2003). In the second case the systems of mobile users perform as interfaces to the Grid computing arrangement that is facilitating job supervision, submission as well as administration of the tasks in an 'anywhere', anytime working and operational mode, as the Grid Computing arrangement offers them with an extensive performance, reliability plus cost efficiency. Physical restrictions of the mobile computing based devices formulate it essential regarding adjustment of the services that Grid is able to offer to the clients' mobile systems. In those situations mobile Grid has the outlines of 'gridifying' the new technology based mobile systems resources. In the subsequent situation of including mobile Grid resources, we should emphasize that the efficiency of present mobile devices those are considerably augmented. PDAs and Laptops are able to offer combined computational potentials when meet in hotspots, establishing a Grid Computing framework on site. According to Litke A. , Dimitrios, & Theodora (2004) this potential is able to help the practice of Grid Computing applications yet in situations where this would be unreal.   Though, there will for all time be the difficulty on why a Grid Computing solution should be assumed in contrast to some other non-Grid support based IT solution (Litke A. , Dimitrios, & Theodora, 2004).

Grid computing technology based arrangement is not anticipated to be the 'panacea' to the issues regarding IT field. It is an assuring rising technology that has the goal to offer more competent as well as valuable solution as compared to its 'competitors' through facilitating the straightforward 'connect and share' method in the similar way as the present Internet search engines implement the 'connect and acquire details' idea. Through this, mobile Grids and Computing Grids those are able to offer perfect solution intended for several huge scale implementations that are of active nature as well as necessitate transparency intended for users. As (Litke A. , Dimitrios, & Theodora (2004) stated grid will augment the job performance and capability of the concerned applications plus will augment operational rate of resources through implementing well-organized means for resource management in the huge amount of its resources. It will facilitate superior forms of supportive work through permitting the faultless incorporation of data, resources, ontologies and services (Litke A. , Dimitrios, & Theodora, 2004).

*1.5- Research Questions*

This study aims to come up with a *best grid computing Model approach for mobile devices among different models* that has good load balance, high processing efficiency, less network communication and thus suitable for mobile computing environment.

1. What is grid computing and mobile computing?

2. Why we need to go for grid computing? Feasibility analysis

3. Mapping of grid computing techniques on mobile domain

4. How grid computing could achieve in mobile devices?

5. What are existing grid computing models for mobile? (information service GC Model, Programming Models, Business Models)

6. Detail analysis of its Design, Implementation, Architecture and Limitations

7. Effect of grid computing on mobile's load balancing, high processing efficiency and less network communication.

*1.6- Objectives of the Research*

In this study we present a model of a grid-based problem-solving environment for wireless mobile devices with limited processing power. Its primary purpose is to allow mobile devices with limited resources to solve problems that they would not be able to solve individually to solve individually. This goal is achieved by redistributing the computational load among many computing devices. This study aims to come up with a grid computing approach for mobile devices that has good load balance, high processing efficiency, less network communication and thus suitable for mobile computing environment.

*1.7- Scope of Research*

This research is going to assess and analyze the distributed computing technology of grid computing. This technology is one of the next generation technology that is promised to offer a great deal of services and support for enhanced corporate and business management. This research is aimed offer a deep and detailed analysis of some of main areas aspects of the grid computing technology. This research will investigate the some of main areas and aspects of the grid computing technology and their fundamental contribution to real life areas. Here I will discuss the structure and contributing of such technologies for the enhanced analysis of grid computing advanced support aspects.

### 1.8- Research Methodology

This research is based on the qualitative research methods. In this scenario, I will perform qualitative research on number of grid computing technology areas for the better analysis and investigation. Here I will make use of academic libraries, databases and websites for analysis and investigation of research martial.

### 1.9- Thesis Structure

The first chapter is about the Introduction of research. The second chapter is about the literature review of grid computing technology. The third chapter is about the methodology used in grid technology development. The chapter four is about the analysis of the grid technology and possible implication. This final chapter investigated the possible results of the technology implication and performance. The last chapter has offered conclusion and recommendation about the technology.

<u>**Chapter 2: Literature Review**</u>

2) **Literature Review**

### *2.1- Grid Computing Technology*

Grid computing technology has been developing since its early foundation in 90's. In past, grid technology was an idea as well as started to have initial accomplishments in corporations and universities. These days, it is a truth that can be employed not simply in scientific or academic atmospheres however in commercial bids to offer leading edge systems to clients. Grid computing technology is based on idea of distributed computing that is taken to the next high-tech evolutionary level. Numerous things have occurred from the binging and technical application of technology based solutions. According to Wang et al (2005) in current years, grid computing was employed in high level corporations, businesses and scientific laboratories as well as attained maturity as the idea turned out to be a model. That model initiated a new means of thinking in the information technology community as well as supported to resolve needs for end users plus the clients of the information technology corporations (Wang et al, 2005). Descriptions similar to desktop grid, server grid, information grid, and lots of there variants appeared in this new age of the technology. The grid computing technology model stimulated numerous technical people external the places where the ideas was born. As Chu & Humphrey (2004) stated it was incorporated into technology based products of together initial and well recognized information technology enterprises. The GC paradigm opened the new means to real pieces of IT that could be employed intended for commercial functions (Chu & Humphrey, 2004). By means of grid computing paradigm, we are able to observe an extremely interesting development: that initiated in researches and in projects guided through universities turned out to be a theoretical architecture that is able to be employed by means of commercial targets to augment a user's experience as well as to facilitate by means of the needs of the corporate that consume IT (Redbooks, 2005).

### *2.2- Developments in Grid Computing*

The idea of the grid computing is not very old idea. The implication of grid computing technology was previously given that grid computing is not basically a novel idea. Chu & Humphrey (2004) stated that it has been under research and under development since lots of pervious years at development organizations and scientific research intended for majority of computing intensive systems. The idea of grid computing is limited to, circuit design, aerospace simulation, and human DNA sequencing based research and analysis projects (Chu & Humphrey, 2004). The idea and applications of parallel processing and server farms are early pioneers to contemporary grid computing technology; though, grid computing potentials are obtainable these days for business utilize based on more powerful support facilitating technologies like that faster processors, network bandwidth and progresses in grid computing technology middleware applications, all at realistic prices to the end users (Wang et al, 2005).

The technology of grid computing is also known as "network computing." As this idea would involve, the technology based areas those deals regarding communications as well as by computing.

As we contract the idea of grid computing with the history of the electrical-power-grid, that covers more than 2 centuries. The idea of computing grid or whole computer technology has a history of fewer than half a century (Gen & Fang, 2009).

Chu & Humphrey (2004) stated the idea of grid computing nowadays recalls the initial days of the Web. At that time technology areas emerged, new and more enhanced web services evolved; things started slowly, however once standards as well as applications emerged as well as coalesced, expansion rapidly ensued (Chu & Humphrey, 2004). These days, the entire technologies of grid computing have been circumstances specific application and operational support. For example if anybody installed the distributed.net user, in this scenario he could online able to access or process work from the SETI@Home grid. In this situation that person will not be able to establish United Devices client solution without making use of their distribution as well as management system (Phan et al, 2003). Luckily, grid standards technologies, structures, open applications, as well as off the shelf software are now emerging quickly. According to Minoli (2004) currently, grid computing technology has instituted to enhance the Web based technology services to identify standard interaction intended for business services similar to business procedures outsourcing (Minoli, 2004).

*2.3- Mobile Grid Computing*

The present grid computing technology structures as well as algorithms are incorporating mobile or wireless computing systems and atmosphere as wireless and mobile technology based devices have not been critically recognized as suitable tool for computing technology resources or interaction in grid computing technology communities. Currently it has been placed attention to incorporate two biggest rising technologies to develop a new and more enhanced technology known as mobile grid computing. For example, in research performed by (Phan, Huang, & Dulan., 2002), (Clarke & Humphrey, 2004); they do not explain on how the wireless or mobile systems can be used present grid computing structures.

According to my assessment mobile grid technology combination offers two enhanced roles in mobile devices in grid. Initially, mobile devices are able to be employed as technology based interfaces to the grid. Therefore, wireless/mobile systems are able to the started as grid based technology resources to observe the tasks being running remotely, as well as to take some outcomes as of the grid computing arrangement.

More interestingly, 2nd aspect of this technology is that: mobile devices are able to be taken as to contribute in grid structure as computing and communication resource providers, not immediately as technology services recipients.

Park, Ko, & Kim (2003) stated there is a great deal of hope that current developments of technologies on wireless and mobile systems and application for collaboration and communications make this new technology paradigm more feasible (Park, Ko, & Kim, 2003).

A major feature of the mobile grid computing technology is the alternating linking of mobile devices. In this scenario, we can discover analogous conditions in Peer-to-Peer computing region. Generally, point to point technology based system composed of enormous amount of computing systems as well as they are able to perform either as a server or a client. In a point to point technology based arrangement, every system's storages, CPU cycles plus contents are able to be shared so as to expand their technology based resource boundaries (Ledlie, Shneidman, Seltzer, & Huth, 2003), (Wilcox-O'Hearn, 2002) & (Berman & Wolski, 1997).

In this scenario, Wilcox-O'Hearn (2002) stated SETI@home project (SETI@home, 2012) offers a flourishing story that huge point to point system are able to be efficiently employed intended for high performance systems through combining processing of desktop computers. In a point to point system, clients are free to leave or join the communication network as well as one research has highlighted that a network communication node remained in a connection state simply for 28 percent of time on average (Wilcox-O'Hearn, 2002). These common connections failure of the network node reduces the computing power considerably and offering the redundancy of workloads is the means employed to balance the shortage (Kondo, Casanova, Wing, & Berman, 2002). Alternatively, the similar work unit is spread into a number of numbers of network communication nodes as well as the initial outcome produced through the majority of consistent and the fast communication node is received. Though, this policy of redundancy is not able to be implemented in mobile grid computing architectures. As the redundancy assures the enhancement of computing performance, it as well wastes the technology based resources significantly. Though the policy is able to be productively implemented in point to point technology based systems where computational resources are rich, comparatively small amount of network communication nodes plus high sensitivity of technology resource dissipation in mobile grid computing arrangements will not permit great number of repetition of work units.

Park et al (2003) have outlined numerous methods to deal with grid computing and technology related performance degradation aspects. Park et al (2003) presented an innovative grid computing based scheduling algorithm that involves the intermittent connectivity of wireless/mobile communication nodes. As grid computing composed of numerous separately controlled local resources (for example clusters, MPPs and workstations). In this scenario, communication scheduling model would fall into 2 groups: local and global scheduling (Park, Ko, & Kim, 2003).

A variety of global scheduling algorithms for grid computing arrangement have been working like that application level cloud computing scheduling (Berman & Wolski, 1997), economy scheduling (Abramson, Giddy, Foster, & Kotler, 2000), high-throughput scheduling (WISC, 2012) and data centric scheduling (Jai-Hoon, Park, & Kim, 2003). Once the grid computing task is planned and presented through global scheduler. It is listed again through local resource managers similar to LSF[1], PBS[2], and Condor (WISC, 2012). FIFO is the well known famous job scheduling strategy in the local managers, however more refined methods and algorithms are offered inside the specific managers.

### 2.4- Mobile Grid Computing Present Research

There are a number of attempts paying attention on developing and designing the Mobile Grid Computing frameworks. Though they are in a beginning stage as well as are predictable to offer the initial results very soon. For example the K*-Grid project[3] is a scheme in Grid computing research that is established through Korean Ministry of ICT. The central objective of the K*-Grid project is to offer a potential research atmosphere to together academia and industries. In this research and development project, a research on mobile grid computing technologies is involved that makes use of inactive technology based resources of a huge number of mobile or wireless systems plus the development of a mobile grid computing based technology platform. As Litke A. , Dimitrios, & Theodora (2004) stated the capacity of the effort comprises the study of wireless/mobile communication networks, technologies and systems as well as the basic needs intended for mobile grid, implementation and design of mobile grid computing structure that foundational on wireless LAN and PDA based technology arrangements (Litke A. , Dimitrios, & Theodora, 2004).

Another example of mobile gird computing project is AKOGRIMO project[4] that is a European financed developed intended to design and develop a blueprint of a Next Generation Grid Computing technology foundational upon the Open Grid Services Architecture (OGSA) that develops as well as intimately co-operates by means of developing Mobile Internet arrangements foundational upon IPv6. The idea of the grid technology based project is to assess the resulting Mobile Grid Computing technology by means of test beds that are selected foundational on obtainable developing applications as of the domain of e-Learning, e-Health and Crisis management (Litke A. , Dimitrios, & Theodora, 2004).

---

[1] http://www.platform.com/products/LSF/
[2] http://www.openpbs.org
[3] http://www.cs.wisc.edu/condor
[4] http://www.akogrimo.org

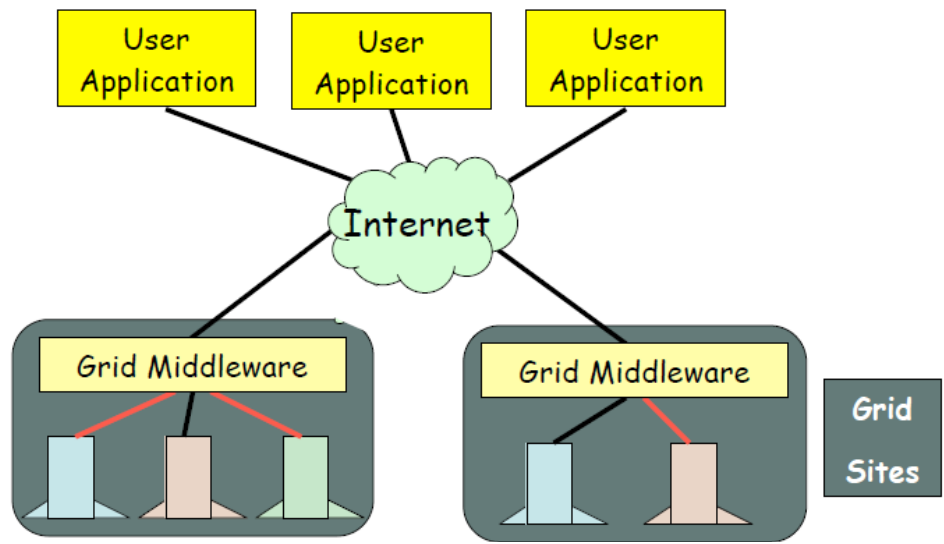

**Figure 3 Architecture**

Source http://inka.htw-berlin.de/wci/05/praes/Prof.Dr.HermannHessling_web.pdf

## Chapter 3: Method

3) **Method**

### 3.1- Mobile Technology in Grid

Mobile Grid Computing paradigm is a new technology based platform that have addressed mobility problems and facilitated both mobile and fixed users to take advantages of enhanced access to both mobile and fixed Grid computing resources perorations clearly and powerfully the fundamental technologies. Mobility engages a group of problems intended for both cooperation and users mode that will be obtainable shortly in the development (Chen et al, 2003). Traveling clients of this technology move around among diverse terminals plus ask session movement (client mobility). For instance, a user is able to open a session on a workstation at his workplace and then inquire its progress to a PC at home. Roaming clients, on the other hand, moving all through by means of their own wireless devices, desired to carry-on their sessions when they cross diverse localities, as it occurs nowadays in mobile phones based tele-communications. In above mentioned both situations the middleware should conceal the particulars of the dynamic use of service mechanism, when the session transfer is necessary, as well it should supervise all the communication session hand-off visibly to the client and perhaps to the application means. As Schulzrinne & Wedlund (2000) one more type of mobility is the communication and collaboration service mobility that is intended as the capability of the structure to uphold the services intended for a client as he is mobile. These support and communication based services should be self-governing as of the systems as well as from the place of the user by means of respect to the communication network user is accessing (Schulzrinne & Wedlund, 2000).

In this scenario, one feature of mobility is that: it should be recognized as difference among mobile systems user interfaces and wireless networking. Wasson et al (2004) stated that in wireless communication based networking arrangement the issue depends in the reality that the association is extremely changeable together in price and quality of systems, plus furthermore that disconnections occur without warning, therefore comprising synchronous communication uncertain(Wasson et al, 2004). In scenario of mobile technology client interfaces the issues approached as of the limitations imposed through the small device size. Small displays are able to simply outline restricted amount of data, as absence of keyboard makes data entry inflexible (Park, Ko, & Kim, 2003). It should be observed that mobility stresses the reality of the synchronous mode of collaboration. In a communication intensive situation the majority of issues are regarding facilitating a synchronous means of collaboration among the peer units. As asynchronous approach is simpler to be established in undependable as well as hard situations the synchronous means brings in a set of restrictions that are through nature hard to manage. According to Litke A. , Dimitrios, & Theodora (2004) the shared items are no more databases and workspaces where numerous users perform on, however moreover methods have to be implemented for facilitating shared systems and parallel schemes for the communication and collaboration session handling (Litke A. , Dimitrios, & Theodora, 2004).

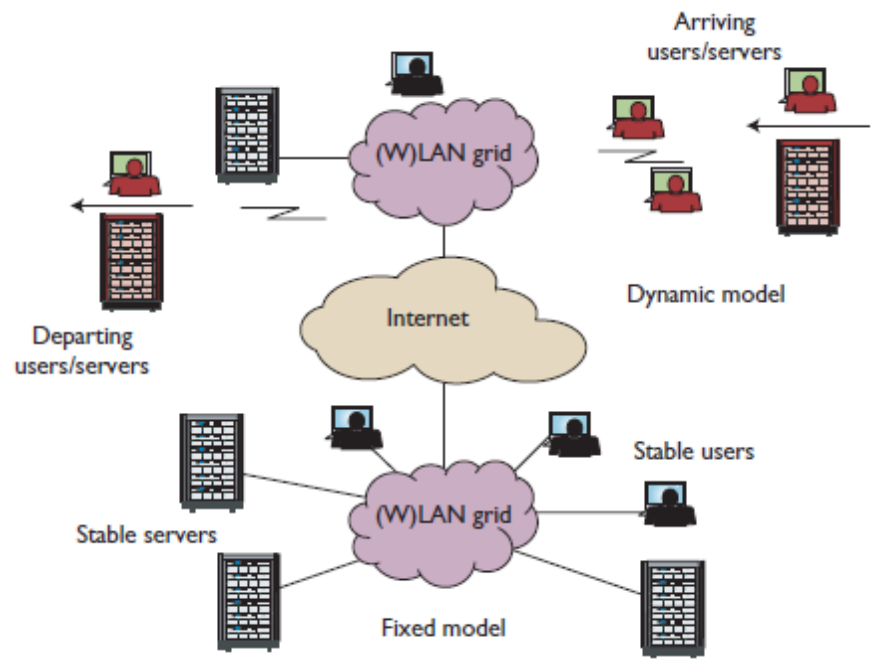

**Figure 4  Dynamic and fixed wireless grids**

Source: http://www.sobco.com/papers/internet.computing.2004.pdf

*3.2- Feasibility Analysis*

The mobile Grid Computing paradigm will initiate transformations to the wide-ranging Grid concept. Fresh and new technology based functionalities of the Grid Computing will be desirable as the old ones will not utilize the entire of potentials that will be accessible. These supporting capabilities will entail end-to-end arrangements through stress on QoS (Quality of Service) and safety, and interoperability problems among the different technologies concerned. As Wasson et al (2004) outlined effective security strategies and methods to address huge scale and mixed settings will be required. In addition, the mobile, unstable and poor networked settings have to be addressed by means of flexible QoS features those have to be contextualized by means of respect to clients as well as their profiles. Mobile Grids will utilize the value that numerous mobile clients distinguish because of the higher potentials of their mobile systems (Wasson et al, 2004). These higher potentials refer to the judgment of today's mobile systems by means of the ones that are present in the past. Though mobile systems are subject to physical restrictions because of their nature they have the capability of having storage and computational potentials analogous to personal computers, high quality user interface, numerous devices communication interfaces, etc. This significance is able to be improved into profits for technology service providers. This involves transformations in different business models as well as policy problems. According to Kuang et al (2002) complex and hard to manage workflows intended for businesses will be required as well as VO or Virtual Organizations will be improved by means of the prospects intended for automatic association plus resource sharing systems (Kuang et al, 2002). Assessments of the different services as well as particularly by means of respect to the Sharing/Service level conformity have to be modified concerning QoS features. Enterprises have to concerns and strategies that will handle the issues among public rights to their clients and feasible models for processes. Furthermore the fair utilize of Grid computing have to be determined through reconciling privileges of public access to technology based resources plus private possession of arrangement plus resources of collaborations and communication (Litke A. , Dimitrios, & Theodora, 2004).

### 3.2- Why we need to go for grid computing?

This section will assess the basic motivation regarding why we need mobile grid structure. Here I will present some of main reasons and aspects those are fundamental behind this research. This subsection of my research thesis comprises number of possible causes and opportunities for business and organizations to look into the grid computing technology as well as consider planning its ultimate deployment for business and corporate use. The major points this part of research will assess is the possible savings those are attained through the grid computing. Chu & Humphrey (2004) assessed that grid computing industry is emerging and developing day by day with rapid pace and more and more people are interested in this new technology based arrangement. The grids based computing arrangements are constant settings that facilitate software systems to put together displays, instruments as well as computational and data resources that are handled through various businesses in extensive corporate areas (Chu & Humphrey, 2004).

According to Phan et al (2003) grid computing based technology atmosphere permits businesses to share databases resources, computing power as well as other tools securely all through the institutional, business as well as geographic limitations without letting down local autonomy (Phan et al, 2003).

Another main motivation that inspired us to chose this technology for mobile networks is that; Grid computing facilitates people from dissimilar organizations as well as locations to collaborate together to resolve a particular issues, like that systems working and operational design collaboration. Grid computing based system development software platforms permits enhanced technology based resource sharing, resource discovery plus association in a distributed atmosphere (Chen et al, 2003). Though grid computing based technology arrangement has been utilized inside the scientific and academic areas for last decade and offering a great deal of support to these working areas . In this scenario, Wang et al (2005) outlined that grid computing technology toolkits, standards, products as well as supporting technologies (for example ubiquitous-broadband-networking arrangement), are now turning out to be obtainable that permit companies to more and more utilize and obtain the benefits of this structure of "outsourced" computing (Wang et al, 2005).

The advantages of grid computing based technology, regarding industrial can be superior productivity attained by better flexibility as well as speed of implementation, all through the enormous computing collaboration, power plus better cost savings (Kuang et al, 2002).

These days approximately all corporations are having main objective is to save Information Technology costs because IT funds are for all the time under pressure because of growing demand for new technologies and services. According to Wang et al (2005) businesses spend more and more budgets 10 percent of their top line profits on IT associated projects, by means of six percent being typical. Service oriented companies tend to be at the superior end of this level. These prices have turn out to be somewhat fixed over the years, as assessed by (Wang et al, 2005).

A novel prospect at the present exists to tackle these price components. A number of these corporate savings are able to be achieved through restoring a technology by means of another. Other corporate savings are able to be attained through technology generalization (for example virtualization of corporate functions).

According to Chu & Humphrey, (2004 ) Technology based service providers that are not intended to invest in areas of information technology and other related utility services will be under growing pressure to demonstrate they are able yet offer cost effective and flexible information technology structure services facilitating constant altering corporation implementations and procedures (Chu & Humphrey, 2004).

In this scenario, Gen & Fang (2009) stated that "Utility Computing will offer support key transformation in system management, architecture, pricing and IT service/product packaging. The corporation investments, though, will be comparatively unpretentious in 2004. This aspect guide vendors to sharpen business marketing message from a marketing cloud to accomplishment road maps intended for company issues. As well, the apparent management in the marketplace could transform" (Gen & Fang, 2009).

According to (Phan et al, 2003), grid computing based virtualization offers greeters infrastructure that is supporting better flexibility and efficiency for corporate areas. It is also assessed that this offers a greater support regarding reduction of overall operational costs up to 50 percent or more. The more detailed cost reduction overview is outlined below:

- IT services deployment costs minimization from 30-80 percent

- Reduction in capacity planning costs up to 5-40 percent

- Reduction in self adapting technologies cost up to 80-100 percent

- Information security related costs reduction up to 20-30 percent

- Reduction in systems usage metering costs up to 4-30 percent

- Reduction in IT systems upgrading as well as migration costs up to 20-40 percent

Noticeably, these figures are tremendously positive and are able to not be realizable in every cases, or even in majority cases. However they do signify a "stake in the ground" concerning finances through vendors (like that HP in this situation).

According to another research regarding advantages by Wang et al (2005): grid computing it is assessed that: Grid computing has facilitated Digex to minimize the time that is required to determine a system performance issues from one hour to immediately 15 minutes. This aspect has reduced the client response time through 75 percent. Grid computing offered a great deal of support regarding accessibility as well as built-in monitoring characteristics to allow junior data center employees to assume jobs for which DBAs were before necessary. Grids computing technology permits to augment the databases managed through every data base administrator from 40 to 68, an augment of 70 percent, and augmented business without growing technology based resources. To emphasize the power of grid computing, one simply requires looking at these days usual corporate computing atmosphere (Wang et al, 2005).

According to a number of researches especially by Chu & Humphrey (2004) the overall cost of IT regarding systems storage capacity, systems processing power and network bandwidth, the volume of a corporation's IT budget is yet associated in processes as well as maintenance process. Corporate administrators are handling and controlling services in the range of 30 to 50 server systems each, as well as these arrangements can be normally utilizing simply a small division (for example up to 5 to 25 percent) of their technology based resources (Chu & Humphrey, 2004). A number of businesses have thousands of separate production-level-servers established for corporate operations. Grid computing based virtualization visibly tackles these problems directly. Grid computing technology is rising as a feasible technology that companies are able to utilize to obtain higher profits as well as productivity out of information technology resources. Grid computing technology is a very much promising technology intended for 3 fundamental reasons (Wang et al, 2005):

- Its capability to formulating more cost efficient utilize of a given amount of technology based resources.

- As a means to resolve issues that are not able to be approached without a major amount of computing potentials.

- The realty the technology based resources of numerous computers are able to be considerately and possibly synergistically harnessed as well as managed in association toward general goals.

The dispute is formulated that what these businesses require is not more hardware, however more proficient utilize of accessible hardware. Businesses require means to tie the entire of these under-utilized machines jointly into a number crunching pool, handle those resources, as well as offer safe and dependable access to virtualized technology based resources (Kuang et al, 2002). According to Minoli (2004), in absence of huge end user retraining or an enormous investment in latest IT structure, promoters of grid computing technology envision an extra useful workforce (Minoli, 2004).

### 3.4- Mapping Of Grid Computing Techniques on Mobile Domain

Mobile grid architectures are able to be extensively categorized into the following 4 types (Ahuja & Myers, 2006) & (Kurdi, Li, & Al-Raweshidy, 2008) foundational on the devices major in the grid as well as the comparative mobility of the systems in the grid.

• Ad-Hoc computing grids

• Fixed-Wireless computing grids

• Sensor Network computing grids

• Mobile/Dynamic wireless computing grids

### 3.4.1- Ad-Hoc Computing Grids

Mobile ad-hoc network composed of systems by means of a extensive level of heterogeneity. These mobile systems vary from comparatively dominant computing devices performed through a vehicle, to extremely minute, low-power communication sensors that are able to be implanted in body of human. Though they can identify small regarding the uniqueness as well as potentials of each other, a group of mobile communication devices are capable to systematize a extremely dynamic as well as adhoc and infrastructure-less network, in that communication nodes are able to correspond in a hop by hop mode. In this scenario Li, L.Sun, & Ifeachor (2005) stated that there is need to put together the technology resource aggregation model of computing grid by means of mobile ad-hoc communication networks, as a result like to construct a mobile adhoc grid computing structures that is able to be instantaneously built at anywhere or any time (Li, L.Sun, & Ifeachor, 2005). Such technology based arrangement have been created as of a group of wireless or mobile systems, an ad-hoc commuting grid would permit the networked systems to achieve a specific assignment that perhaps away from an personal communication or computing capability. Instances of systems of mobile ad-hoc grids can be wild fire fighting, disaster management as well as e-healthcare crisis management, etc. The wide-ranging features of mobile ad-hoc communication networks are able to be enhanced to ad-hoc computing grids as well as therefore ad-hoc grids are credited regarding utilize of power and bandwidth limitations, communication network partitioning, multi-hop message delivery as well as communications randomness.

### 3.4.2- Mobile/ Wireless-Sensor Network Computing Grids

Mobile-Sensor networks are building through small devices that are normally devoted to a particular function. By means of every system in the grid is devoted to a particular function, the grid itself can be contains numerous dissimilar forms of systems so as to achieve the objectives of the specific computing grid. As stated in (Xing, Lu, Pless, & Huang, 2004), mobile or wireless sensor networks puts together processing, detection as well as communication into the computing grid. The communication sensors getting part in a sensor grid can be still once they have been positioned (Miller. & Vaidya, 2004) or they can be moving or mobile. Mobile/Sensor networks are at present utilized for the sake of checking ecological aspects like that humidity or temperature transforms, light and motion power. The progresses in sensor networks are taking toward progresses in physical building security, agriculture, industrial and fire-fighting areas. Mobile/Wireless sensor network that are able to be positioned in commercial and residential structures to check human attendance as well as turn off the lights when no citizens are noticed (Kanellos, 2005), is under development in California by Dust Networks of Berkeley. RFID computing technology systems are used to track products as of manufacture in the course of delivery and distribution plus possibly, to the customers shopping-cart leaving the shopping store (Gilbert, 2005), is developed.

### 3.4.3- Fixed Wireless grids

The mobile/wireless grid computing expands grid resources to wireless systems of different sizes as well as potentials like that laptops, sensors, edge devices and special tools where these systems are typically static. In mobile/wireless grids computing arrangement, mobile or wireless devices are able to perform as actual grid based communicating nodes where element of data processing as well as storage is happening. In a particular kind of wireless grid computing arrangement, the entire types of wireless systems are recognized with pure access devices with no system storage or processing potentials (Foster & Kesselman, 1997); necessary technology resources are taken from a wired backbone grid with resource rich support. A lot of technical issues occur when incorporating wireless devices into a wireless grid computing structure. These comprise low bandwidth as well as high security issues, latency and power based resource consumption. As a result, numerous communities, comprising the Interdisciplinary Wireless Grid Team, there is such platform for the community that is accessible at wireless grids.net. This web based wireless grid computing community is exploring these fresh problems and make sure that future grid computing peers are able to be wireless devices.

### 3.4.4- Dynamic or Mobile Wireless Grids

According to Grimshaw, Ferrari, Knabe, & Humphrey (1999) mobile grids computing structure make grid services available by means of mobile devices for example smart phones and PDAs. As the system's processing power as well as other potentials of these mobile devices augments, commercial corporations and researchers are finding out innovative methods utilize as well as share their technology based resources. When the huge amount of accessible wireless devices is recognized, the power of these dynamic ad-hoc links turns out to be vast (Grimshaw, Ferrari, Knabe, & Humphrey, 1999). By means of wireless grid technology based arrangement, these systems are capable to attach to the Internet, offers peer to peer communication, take benefit of the technology based resources of wired backbone grid as well as formulate their own technology based resources accessible to the wired grids (Baker, Buyya, & Laforenza, 2002) & (Chu & Humphrey, 2004). In state of crisis, for example in state of natural disasters or battle fields, mobile/wireless devices might be the only accessible collaboration and communication systems. The mobile devices incorporated into grid systems are able to dynamically contribute plus offer data services or computational services (Tuecke, Foster, & Kesselman, 2001) & (Waldburger & Burkhard, 2006).

These mobile communication devices as well offer support as an interface to a stationary grid intended for transmitting communication requests as well as receiving outcomes. Occasionally this method is known as mobile access to grid arrangement or just mobile access grids. Newly, researchers have formulated many attempts to implement mobile technology based grids. In this scenario, researchers have projected a variety of methods for establishment of mobile grid idea, comprising central as well as P2P structure with mobile grid middleware, intelligent mobile agents as well as lots of more tools (Tuecke, et al., 2003) & (Tuecke, et al., 2003). Present mobile grid projects comprise Akogrimo (currently accessible at mobilegrids.org) and MADAM (accessible at website www.intermedia.uio.no/display/madam/Home).

## Chapter 4: Analysis

4) **Analysis**

### *4.1- Grid Technology Structure*

The Grid Computing Technology is speedily developing in both directions as well as application areas plus there is a equivalent confusion and excitement as to the "right" means to think regarding Grid Computing technology based systems. According to Gridforum (2012) GCE or Grid Computing Environments approximately explained as the "client aspect " of a computing arrangement that is demonstrated in the image given below where there is a fuzzy separation among grid computing environments as well as what is known as "Core" Grid in the image. The latter would comprise access to the technology based resources, supervision of plus communication among them, safety as well as other similar potentials. The latest idea of the Open Grid Services Architecture (OGSA) (Gridforum, 2012) (that is itself rising) explains the these "Core" potentials as well as the Globus project (Globus, 2012) is the most excellent example of that is recognized as the "Core" software project.

Grid Computing Environment is also described as a group of technologies and tools that permit clients to offer "simple" access to Grid's technology based resources as well systems. Often it emerges to the user like a web portal that offers the user interface to a multi-tier computing grid application building stack, however it can as well be as straightforward as a Grid Computing Shell that permits a user access to plus manages over Grid Computing technology resources in the similar means a conventional shell permits the client access to the file system as well as procedures space of a regular OS. As Fox, Pierce, Gannon, & Thomas (2002) stated that a number of researches assessed that Grid makes use of different systems and technology. The below given image shows the multi tire arrangement of the GRID technology based structure (Fox, Pierce, Gannon, & Thomas, 2002).

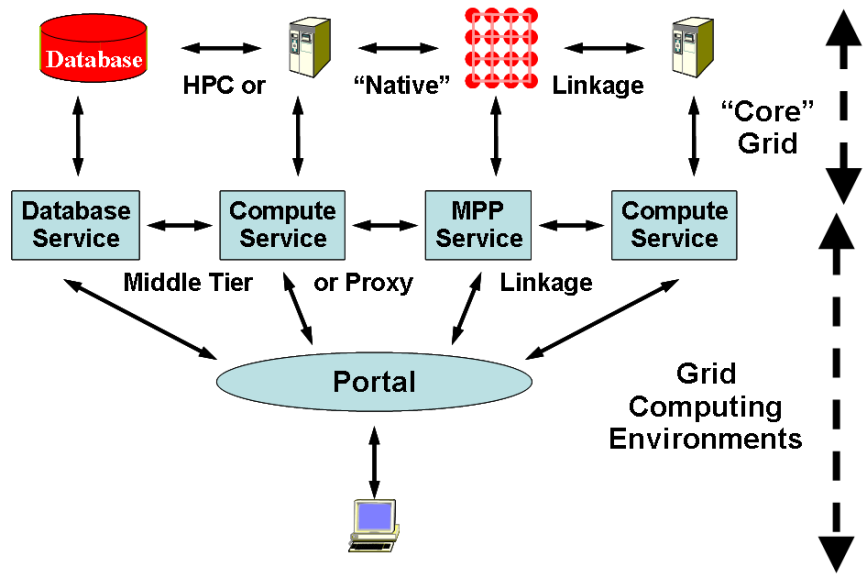

**Figure 5 Middle-Tier of the Grid**

Source: (Fox, Pierce, Gannon, & Thomas, 2002)

As I explained above, Grid Computing Environments accomplishes 2 fundamental functions (Fox, Pierce, Gannon, & Thomas, 2002):

- Client side of the Grid Computing arrangement involves programming.

- The next function is about the managing and controlling user interaction.

GCEs can be categorized in numerous aspects. One simple categorization is about technology based aspects. The different technology based projects be different in scenario of nature of behavior of objects, languages used, utilize of some technology similar to Globus toolkit, Java servlets, or GridFTP as well as other application concerns. A number of these problems are significant intended for architecture or performance however frequently is able to look to the user as not so significant. For example, there is a tendency to utilize more greatly XML, Java and Web Services however this will simply be attractive if the resulting systems have significant characteristics like that improved sustainability, customizability plus simplicity of utilize without giving-up too much in regions similar to technical performance. According to Fox, Pierce, Gannon, & Thomas (2002), the simplicity of development by means of contemporary technologies frequently produces superior functionality in the grid computing environment for a given amount of accomplishment effort. Technology dissimilarity in technology based projects are significant however more attractive at this phase are the dissimilarity in potentials as well as the model of computing implicit or explicit in the grid computing environment (Fox, Pierce, Gannon, & Thomas, 2002).

All grid computing environment systems suppose there are a number of backend remote resources as well as attempt to offer suitable access to their potentials. This shows one requires a number of type of model intended for computing. At the easiest level this is running a task that before have insignificant outcomes as data typically requires being correctly established, as well as access is necessary for executing job status plus final results. More complex instances necessitate synchronized collection of data, numerous models (moreover connected at a known time or following each-other), examination of outcomes, visualization, etc. A number of these events necessitate considerable association among researchers as well as sharing of outcomes as well as designs are required. These guides to the idea of computing environment co-laboratories those are facilitating for sharing of data and information between scientific teams operational on the similar problem area (Fox, Pierce, Gannon, & Thomas, 2002).

We are able to construct a model of diverse grid computing environment methods through analyzing the issues as several kind of generalization of the job of computing on a single system. Consequently we are able to emphasize the following categories of characteristics (Fox, Pierce, Gannon, & Thomas, 2002):

1) Management of fundamental parts of a distributed computing arrangement for example computer plus data resources, files, accounts and programs. The grid computing environment will normally interface by means of a setting similar to Globus or a set scheduler similar to PBS to really manage the backend technology based available resources. Though the grid computing environment will outline the client interfaces to manage these resources. This interaction is able to be straightforward or difficult as well as frequently assembled hierarchically to reveal tools constructed in similar style.

2) In the above given figure shows a three tier model, that is normally used for the majority of GRID systems, shows that some given potential are able to emerge at numerous levels. Possibly there is a backend parallel technology based operational running an MPI task. This is front ended possibly as a service through a number of middle tier parts running on a completely dissimilar system that could even be in a dissimilar network security domain. Anyone can "communicate" through this service at each level; an enhanced performance input output transfer at the parallel computing arrangement level or through a less fast middle tier communication protocol similar to SOAP at the computing service level. These 2 communication calls (part communications) are able to signify dissimilar processes or the middle tier call are able to be coupled by means of a high performance parallelism; normally obtainable in middle tier offers control as well as the back end support for the enhanced raw data transfer. The consequential rather complex structure as shown in above image. We have every element (technical service) demonstrated in together middle as well as HPC network tiers.

3) Most significant and wide ranging characteristic is network security (verification, permission and privacy); those need to be tackled in a number of means or other through fundamentally in the entire environments.

4) Data management is one more generally significant theme that gets even more significant aspects on a distributed system as compared to single systems. It comprises databases, file manipulation as well as access to raw signals as of systems like that accelerators and satellites.

5) One aspect increase the fundamental GCEShell by means of a library of other wide-ranging purpose of tools as well as this is able to be supported through the GCE. Like that systems comprise Grid MPI, Grid FTP, parameter sweep as well as more wide-ranging workflow, plus the symphony of GCEShell primitives.

6) A number of other higher-level systems and methods are as well significant and numerous tend to be rather application dependent; smart corporate decision support and visualization of data elements trough different kind of algorithm to utilize are able to be put here.

7) Through seeing at corporate commercial portals, one discovers that they typically facilitate refined user interfaces by means of numerous sub-windows combined in the user interaction. For example the Apache Jet-speed development is a famous toolkit facilitation this type of computing operations (Jakarta, 2012). This client interface aggregation is frequently facilitated through a grid computing environment.

 In addition to some specific characteristics, a grid computing environment typically involves a particular computing model intended for the Grid Computing as well as this model is demonstrated in the grid computing environment structural as well as the view of the Grid demonstrated to the user. For instance object models intended for systems are extremely well-known as well as this object view is revealed in the analysis of the Grid obtainable to the user through the grid computing environment (Fox, Pierce, Gannon, & Thomas, 2002).

### *4.2- Mobile Grids in Context*

Wireless/Mobile grids computing evolved from a set of the increase of fresh spectrum marketplace business models. This evolution offered modern technologies establishment in different Wireless/Mobile networks, as well as 3 associated computing models: P2P computing, grid computing and Web services (McKnight, Howison, & Bradner, 2004).

The technologies of wired grids are normally aggregations of fixed resources among identified organizations, be they corporate or academic, in better security plus relatively static situations. To contribute in the Grid computing structure by means of the present Globus software, for instance, a machine have to produce as well as preregister an X.509 technology based certificate. These trusted, static settings stand in stark distinction to the circumstances facing the Wireless/Mobile grid computing arrangement. Foster and Imanichi (2004) illustrated the wired grid computing that spotlighted as "mixing of considerable technology based resources to convey nontrivial qualities of IT service inside a situation of at least partial trust (Imanitchi & Foster, 2003).

However, the objectives of grid computing technology are the similar as those that motivate our efforts regarding establishment of secure, flexible as well as coordinated resource sharing arrangement between dynamic sets of institutions, people as well as resources (McKnight, Howison, & Bradner, 2004).

Point to Point networks arrangement, for example Gnutella, Napster and Kazaa, have features in common through Wireless/Mobile grids computing. They have to put together synchronized sharing between heterogeneous devices, all through undependable network communication connections, by means of small or no prearrangement with little or no warning of website failure. According to McKnight,

Howison, & Bradner (2004) grid computing had to produce a constant and standard technology based service structure, P2P arrangement have to address technology based resource sharing in the face of undependable networks as well as end systems (McKnight, Howison, & Bradner, 2004).

In spite of being so overhyped, mainly web based services basically ease remote access to technology based resources. The gird computing structure establishment plan is about the offering access to a collection of technology based services present among a client or provider or among 2 companies. However the new technologies at the center of web based services are lightweight XML, SOAP, message passing, as well as the elastic Web Service Definition Language (or WSDL) implement to some ad-hoc sharing circumstances (McKnight, Howison, & Bradner, 2004).

According to McKnight, Howison, & Bradner (2004) the Wireless/Mobile grid computing technology therefore illustrates on in any case of the computing models at present experiencing fast development. The below given image puts Wireless/Mobile grids in framework that demonstrates how different technologies cover the technical methods and problems of grid computing, web services, mobile commerce, P2P systems, spectrum management and ad hoc networking. This picture demonstrates how different technologies like mesh and sensor networks will eventually cooperate through software radio as well as other systems to resolve Wireless/Mobile grid computing issues necessitates plenty of additional research, however the below given image in any case confines several of the most important facets of a Wireless/Mobile grid computing technology (McKnight, Howison, & Bradner, 2004).

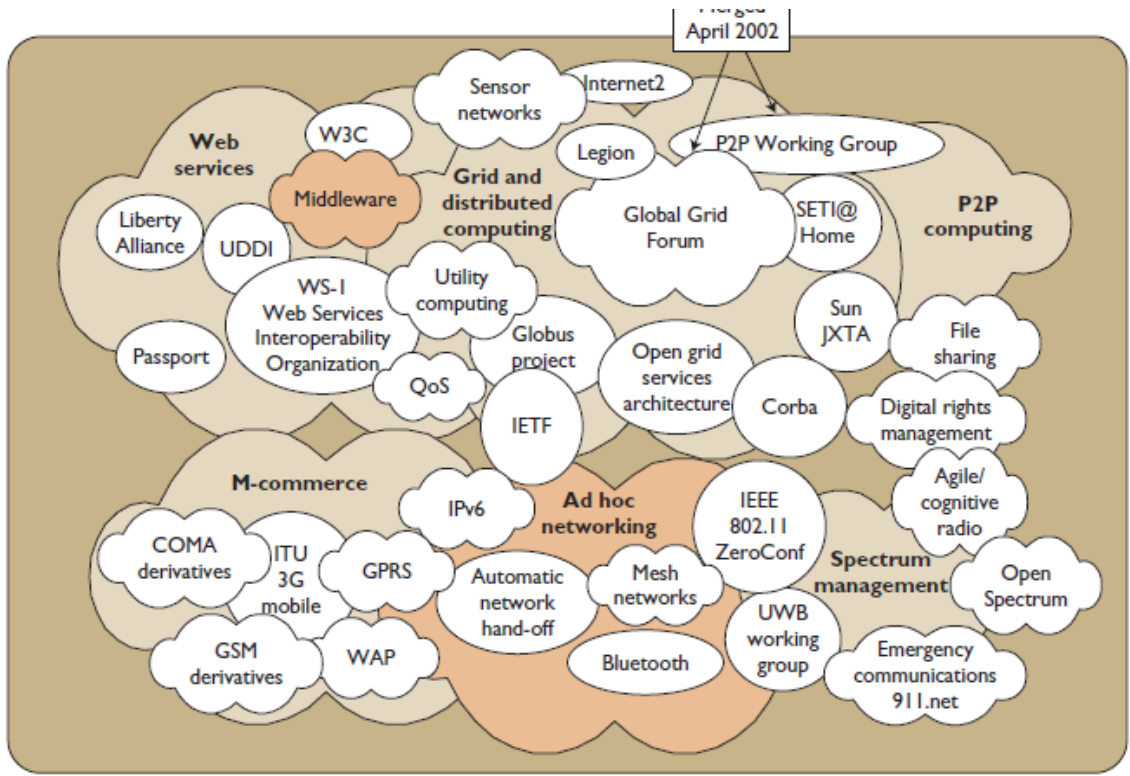

**Figure 6 A wireless grid issues and standards map**

### 4.3- How grid computing could achieve in mobile devices?

According to Kumar & Quresh (2008) the establishment of the mobile technology based services entails two basic issues domains: first issue us about the computational structure intended for the accomplishment of computing services, that looks upon the software and hardware formations that supports the linking of mobile systems (information transport system, devices and others) as well as service base built-in to this situation (fault tolerance based computing services, position recognition service, etc) (Kumar & Quresh, 2008). The second main issues is about the service application itself, that involves developer challenges regarding combining the accessible computational arrangements as well as produce a way-out to the issues of the states in matter. Also as stated by (Guigere, 2012), mobile computing technology enhances a great deal of complexity and inherent to the computing situation, like that mobility, dynamic environments, latency, computing technology resource limitations plus instabilities in information transfer, energy supply issues as well as output/input interface restrictions. Numerous researches are performed as well as methods are being developed to avoid these issues. Among these issues, we are able to mention progresses in the regions of

- Data communication and collaboration, for example reliable and faster networks arrangement (Henricksen & A.Rakotonirainy, 2001).

- Offering support for the distributed computing, by means of processing distribution between server and client

- Software Engineering aspects those involve the formation of applications that adjust their performance as of resource limitations (Guigere, 2012).

Kumar & Quresh (2008) highlighted that though, till now no proposed methodologies have offered complete and comprehensive solution, as well as in this mode they direct to establishment of ad-hoc computing and communication methods, related to problem scenario being dealt. In case of lack of homogenization and re-usability reasons supplementary costs in the building such systems (Kumar & Quresh, 2008).

For the sake of connecting mobile computing and grid computing systems and technology there is need for some additional platform of arrangement. We can call it as middleware intended for the getting a combined support and communication among mobile and grid operational services. Kumar & Qureshi (2008) stated that this middleware technology incorporates application, methods and tools of grid computing arrangement like directory, data communication, operational security, service distribution, technology based resource allocation and resource discovery. These all resources offering systems developer support for the homogeneous and reusable resources handling (Kumar & Quresh, 2008).

### 4.4- Grid computing and Mobile computing Integration

As I previously outlined a combined setup for the establishment of Mobile Grid Computing arrangement, based on that setup, here I will outline the combined grid and computing architecture. This arrangement is having main characteristics of the essential computational power and ad-hoc communications support.

For the sake of application of such arrangement we are having main requirements those are outlined below (Kumar & Quresh, 2008):

- For the establishment of the Mobile Grid Computing arrangement we need a data communication means. These data communication methods are essential for data transmission as of the mobile sensors systems plus incorporation of the personal and actuators assistants.

- The next maim requirement for the establishment of the Mobile Grid Computing technology is the security methods. Fir the application of such huge arrangement security methods are one of main need for better and secure data transmission as well as for access control to service technology based resources.

- Mobile Grid Computing arrangement also requires better methods handling and management of the technology based resource and their registration that permits intended for personal and sensors assistants exiting and entering control in the grid system.

- Another main requirement for Mobile Grid Computing arrangement is the better offering the device management techniques that permits to handle mobile actuators, sensors as well as personal assistants in a better and centralized mode. In this scenario, in spite of environment dynamism, distribution and device heterogeneity, we need to offer better device management means for the enhanced handling and management of technology based arrangement.

- Mobile Grid Computing arrangement also requires better communication and collaboration interface technique for mobile devices that permits the communication by the users or linked systems.

- For Mobile Grid Computing there are some techniques required for the for work coordination among mobile users that permits cooperation among personal assistants.

### 4.5- Application of New Architecture

For the sake of developing a Mobile Grid computing technology based arrangement, there is need for above stated requirements. These requirements are beyond the scope of programming languages (for example

Java-2-MicroEdition) as well as necessitates a software systems that attains device development, integration, reuse, homogenization, synchronization, resolution of tackling issues, resource discovery, safety, as well as other qualities.

It is considered that there exists an association among ways offers through Grid Computing plus the development of systems intended for the states of mobile computing. For example mobile computing needs systems intended for service detection in an active situation organized through data communication channels among systems and technology based service offers. According to Kumar & Quresh (2008), such technology based development tools permits us regarding incorporation of devices as well as distributed technology based resources, as well as security, for example privacy of information plus verification (Kumar & Quresh, 2008).

### 4.5.1- Existing Grid Computing Models for Mobile

This section will discuss the some of main existing grid computing models for the mobile technology. I this scenario, I will outline some of proposed technology based arrangement for mobile grid technology. For example the new grid computing technology         presents facility to establish mobile communication services intended for automated decision support applications. There is numerous application packages intended for application of grid computing technology.

According to Kumar & Quresh (2008), the establishment of decision support services through mobile technology based arrangement requires support of: user interface, collaboration, communication framework awareness, as well as an interaction procedure that permits the building of application beyond the simply reactive. In an atmosphere of partial technology based resources, systems have to be responsive of the technology based computational resources boundaries non-dependable plus irregular collaboration networks, mobility, energy supply boundaries, changing computing environments,  plus  minimized user interaction. These restrictions are natural to  the situation as well as will be pacify and not outdated, through potential communication technological developments. According to previously stated requirements, it is assessed that the present grid middleware technology regarding communication and collaboration support; technology based resource sharing support; context awareness support and execution on mobile systems support, these all main areas need to be managed and handled for the enhanced performance and functionality (Kumar & Quresh, 2008).

Another mobile grid technology based model Globus (Globus, 2012) is also well known in the domain. This model is an open source software package build through the Glob Alliance that presents an application development as well as grid computing arrangement development toolkit. It facilitates intended for mobile communication service that involves the characteristics those discovered in collaboration in the course of resource layer protocols that can be attain information as well as promoting communication, control jobs plus technology based resource distribution. As well there is computing resource sharing offered by means of the resource manager that is also known as the Globus Resource Allocation Manager or simply GRAM. This resource manager offers job transmission and monitoring interface for the mobile grid computing arrangements.

Another Mobile Grid computing model is Gridbus[5]. Gridbus is developed through the GRIDS or Grid Computing and Distributed Systems Laboratory. GRID is established at University of Melbourne. GRIDBUS is an open source grid software package intended for building technology based mobile grid technologies for online Business and Science.

In case application of mobile grid computing technology there is used of numerous other middle ware systems like that Unicore, Globus, Alchemi [6], and others. Its facilitates intended for mobile services characteristics that is available through communication and collaboration, dynamic  settings and resource allocation offered through lower level middleware or foundational middleware technologies like stated earlier.

Another mobile grid computing model for middle ware technology is known as Legion. Legion is a middleware technology based framework that build through a project at   University of Virginia, as well as is described as a meta system foundational upon the objects (grid technology resources) by means of billions of hosts plus millions of objects associated jointly through high-speed communication networks, super-computers in a system, workstations those are able to combined unusual structures, physical locations as well as operating systems. The technology based support offered through Legion intended for mobile communication and collaboration service characteristics is observed that binding system support probable association by means of Tuples like that plus management (LOA) and supply object addressing (LOID). There as well exists technology based resource sharing offered through Legion Object Addresses (or LOA) that includes a physical address since the Internet Protocol and are able to distribute these technology based resources by means of multicast k (Kumar & Quresh, 2008).

Another one of main Mobile Grid Computing arrangement is UNICORE that is also known as Uniform Interface to Computer Resources  is a grid middleware arrangement that puts together technology based grid computing resources by means of the a graphical interface developed in Java programming language. It facilitates for mobile service features those are able to be discovered in collaboration offered by means of UNICORE online servers following the user/client verification. The association is attained by means of the online resource servers by transmitting tasks to peer UniCore gateways that performs the task plus reacts. The online web based technology resource support and provision is attained through the AJO or Abstract Job Object that is an object that transmits as well as receives tasks (Kumar & Quresh, 2008).

***Analysis of Design, Implementation***

This section is about the analysis of the design, implementation, architecture and its limitations of the mobile grid computing architecture. Pervious sections has outlined the motivation and requirements needed for application of the mobile grid computing services. Here I have also outline the essential characteristics of a mobile grid computing arrangement.

**4.5.2- System Design and Architecture**

This section will outline the proposed mobile grid computing technology system design and architecture. This section will present fundamental components as well as communications aspects of proposed system. Here we will integrate the architecture of grid computing with the mobile devices in computational grids.

Below given Figure no. 1 demonstrates the proposed architecture of mobile grid computing arrangement. Here we are having a layered approach among system components for communication and collaboration.

- Proposed architecture of mobile grid computing is having a middle ware section that serves as a connection that connects
- In the next section we are having applications and systems running on mobile devices as well as the grid computing support and collaborative services operational on central level of the grid administration offices.
- The third level is about the central offices those are having Directory Module stores service and data in S1 communication network directory databases.
- In the proposed arrangement next section is the Broker Module that handles the technology based schedules and resources tasks plus checks their running, monitoring, storing operational data as well as the tasks outcomes in S2 task directory databases.

To convince the basic mobile grid computing system basic requirements those are listed in pervious section, we need to consider the system's architecture. To better manage and handle the system's operations following main procedures are need to be followed:

I. To convince the basic needs and requirements of mobile grid computing architecture's data communication techniques, the HTTP protocol is employed, assuring compatibility among devices.

II. To better suit the system's stated requirements of system's security in data transmission as well as access control management, verification techniques as well as cryptography are employed in network communication node registrations.

III. To suit grid technology based resource registration, the Directory Module makes use of the LDAP protocol as well as upholds the gird computing resource registrations.

IV. To manage the grid computing architecture device coordination and management, the Broker Module directs the devices by means of its supervision techniques, those will be used to satisfy the grid computing technology client interface by means of the communication devices. In this scenario XML technology is utilized for data illustration as well as visualization at the systems.

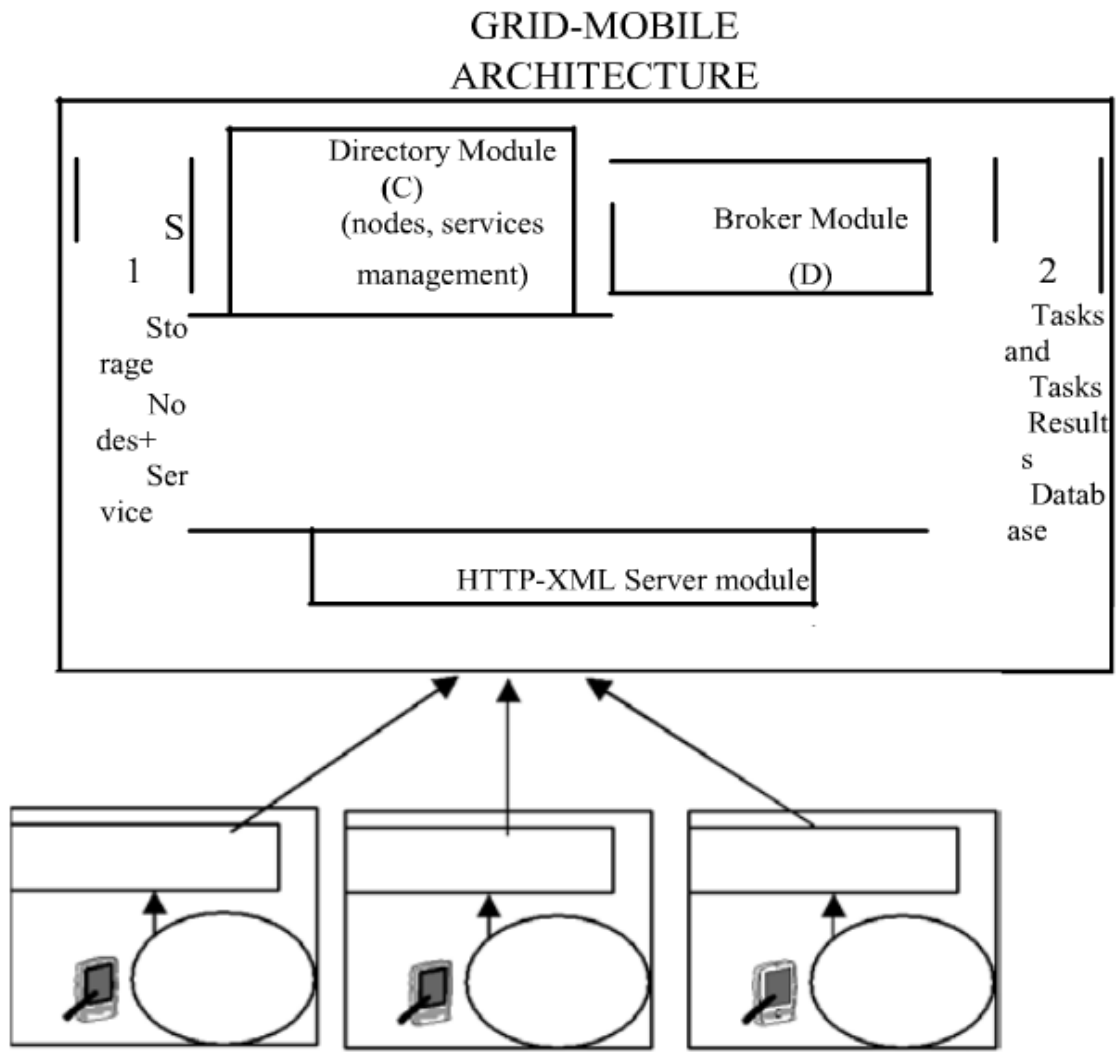

**Figure 7 Grid-Mobile Architecture Source: (Kumar & Quresh, 2008).**

### 4.5.3- Implementation Case Study

This section is about the application of the mobile grid computing technology to specific areas. In this scenario this section will elaborate the case scenario regarding application of proposed mobile grid architecture. Here we are going to apply the proposed grid arrangement for a doctor consults scenario, where a patient healthcare data is being monitored through a mobile sensor system. The mobile sensor occasionally transmits patient's fundamental information for the storage that formulated accessible to medical staff for better assessment of healthcare situation of the patients.

To make such system work we need to take following given steps (Kumar & Quresh, 2008):

- Medical Staff's Personal Digital Assistants or PDAs reorganization and registration on the grid architecture.

- Previous grid network based mobile sensor registration on the network

- Data demands through the hospital medical staff or doctors

- Patient records consulting and search of the grid sensor attached to the mobile network

- Patient data and information transmitted to medical staff and doctor's personal digital assistants.

The below given techniques as well as structural design modules are concerned in the below given described situation (Kumar & Quresh, 2008):

- In step (i & ii) processes intended for validating the personal digital assistants on the mobile grid is to register Node (PDA-type) as well as their local Module runs it.

- In step (iii, iv) the technique intended for data demand is placed job (assess Patient temperature) that responses the against the patient-ID by means of his features as well as sensor potentials.

- In step (v) of the mobile grid application involves a method regarding obtaining Task Results (or taskid) transmits the information as well as jobs into personal digital assistant that requested it.

In the above stated situation, we can confirm that the mobile grid technology based resource homogenization is a major benefit intended for the enhanced administration of the situation. In this scenario, being every grid technology based resource involves a grid   node that is having a middleware.

Through its techniques and capability to handle data and information as well as physical gird computing resources (Kumar & Quresh, 2008) .

The below given Figure no. 2 shows a number of its API processes that facilitate its communication by means of the setting.

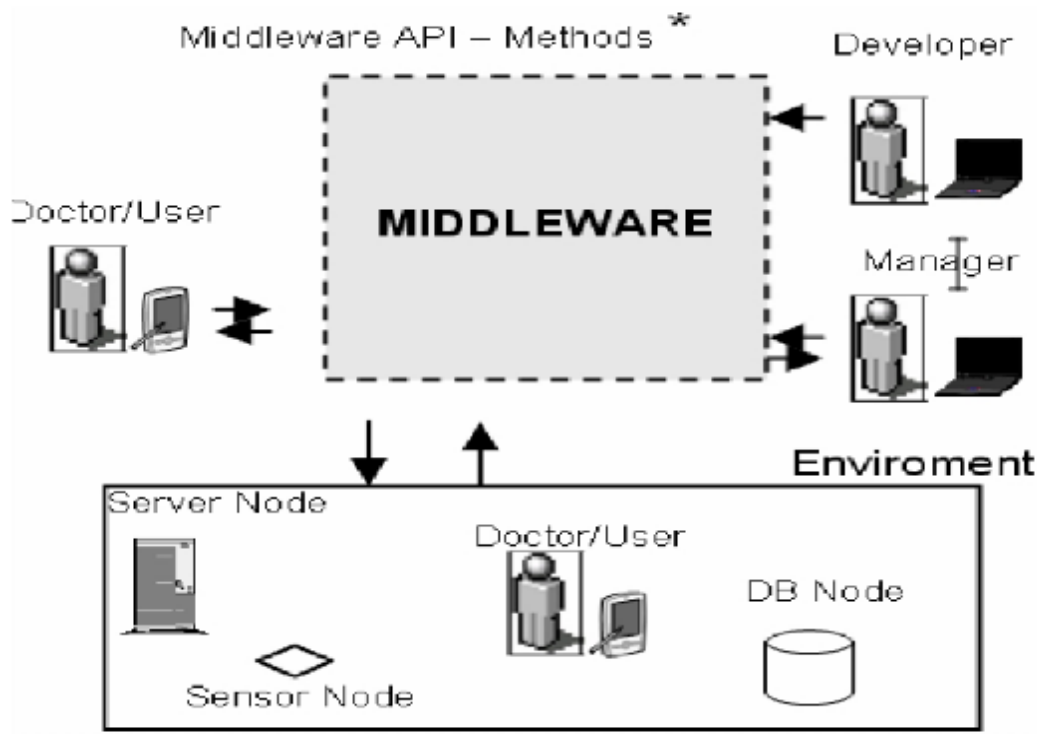

**Figure 8 Middleware API and interactions**

Source: (Kumar & Quresh, 2008).

Above given research based work regarding the establishment of mobile grid computing architecture show all possible aspects of the development of the mobile gird technology in which its fundamental modules is middleware. This middle ware is the interaction among the mobile and gird technology framework. This framework also offers a great deal of the value to the possible aspects of the mobility in grids as well as offers a great deal of homogeneous grid resource handling and administration. This new proposed architecture offers enhanced support for the better handling and management of requirements of management, mobility, safety, data communication and greater interface that taken as the fundamental needs for the establishment of environment for gird computing through offering features of the context awareness, mobility, as well as some time/anywhere data and information support (Kumar & Quresh, 2008).

## **Chapter 5: Results**

**5)  Results**

### *5.1- Mobile Grid Services*

These days mobile gird technology can be applicable in variety of areas of life: For example mobile grid technology can be used for military services. Such mobile grid technology based arrangement will offer a very secure atmosphere for communication and collaboration in battle filed.

Mobile grid technology can also be very useful for disaster management areas. In case of any natural or other disaster we can establish a quick and more secure communication arrangement that can offer a great deal of support and capability for communication and collaboration.

According to Gen & Miao-long (2008) in case of mobile grid technology based structure application, battery power in the most significant aspects of the every grid devices.  So, mobile grid individual devices are likely to lack enhanced reliability.   In this scenario, battery power collapse or intended power management (manual or automatic system power off) augments the failure probability.  To tackle this issue, Mobile OGSI.NET permits a great deal of services to transfer from communication network host to host. Particularly, for better mobile grid performance we need establish fundamental Grid Service state loading and saving.  A Grid Service can save its present condition that another Mobile Grid Service of the similar kind is able to then load as well as carry on running.  For instance MyCounter-Service-291 of kind Counter Service that can save its state by means of Save-Counter-State().  After that one more My-Counter-Service-18 as well of kind mobile grid Counter-Service that takes  the present grid state object by means of Load-Counter-State() as well as carries on running as if continuing from the earlier save point. The fundamental move process requires numerous enhancements (Gen & Miao-long, 2008).

### *5.2- Load balancing, high processing efficiency and less network communication Aspects*

This section is about effects of grid computing on mobile's load balancing, high processing efficiency and less network communication. Here I will assess and analyze some of main areas and aspects of the influences of grid computing technology possible influences regarding mobile communication and collaboration load balancing. Here I will assess the possible influence regarding high processing efficiency and enhancing network communication. A latest mobile computing model intended for data and information transmission services can be implemented through the grid computing environment. Fundamental technologies those can be used in this technology based arrangement are: grid computing, mobile agent based mobile computing, GPS, location based service (or LBS), and wireless technologies based services for communication. These wireless communication support and services can be implemented through transfer control protocol/internet protocol, common packet radio service (or simply GPRS), as well as through UDP or user datagram protocol. So as to manage a less effective network bandwidth as well as instability of the wireless communication based internet, global distributed corporations are trying to incorporate a more enhanced technology based framework that is having a remarkable support for all stated limitation. In this scenario, a latest mobile grid computing agent based computing model that support MSISG (or mobile spatial information service grid) is further projected that has high quality to support and manage the load balance issues of mobile network, less network communication, high processing competence plus consequently appropriate intended for mobile computing situation  (Gen & Miao-long, 2008).

Through application of Mobile computing technology support for Grid Computing arraignment, we are able to get very effective support for enhanced information transmission service and also information access for public through anytime and anywhere access support. Therefore mobile grid computing technology turns out to be part of people's every day life. In this scenario, numerous foundational systems and technologies are currently under research and assessment for better application and handling of the mobile grid computing services and support. In this scenario, we are having grid computing, mobile agents, spatial data model, GPS, LBS, TCP/IP, UDP and CDMA based systems support (Gen & Miao-long, 2008).

Through application of mobile technology based agents in distributed computing architecture resolves the issues inadequate communication bandwidth as well as less steady connectivity faced through the computing applications, inside a mobile computing environment (Kuang et al, 2002). There are numerous high-quality causes of using mobile agents in the grid based computing and communication system (Lange & M.Oshima). Hong et al (2001) outlined that though application of integrating distributed Web based geographical information system via mobile agent as well as GML technology makes simpler the arrangement of clients as well as minimize load of communication network, as well as manages heterogeneity (Ji-hong, Shui-geng, Fu-ling, & Ling-kui, 2001.).

A number of enhanced versions of the Grid based geographical information system prototypes similar to its incorporation by means of mobile agent technology as well as implementing indexing system intended for spatial data ethnology based resources enhances the methods of spatial data detection as well as incorporation (Zhao, Fang, Chen, Yan, & Sun). Gen (2010) stated that mobile communication agent for real-time access and downloading support of spatial data and information in Grid computing based geographical information system atmosphere outcomes in a high-quality load balance support. These technologies also enhance high processing efficiency as well as led to less network communication (Gen T. , 2010).

Incorporating mobile agent in Grid supported information system resolves issues like the lack of network management mechanisms to get the stability of system load, lack of interoperability among communicating system as well as lack of asynchronous computing methods of customary information systems, those execute in relative low operational performance position. According to Tong & Shen (2009) though application of such technology based support we are able to make comprehensive utilize of enormous distributed computing resources, as well as we are able to offer a great deal of flexible services to clients (Tong & Shen, 2009).

Mobile Grid computing technologies and systems facilitates in resolving issues regarding different information systems interoperability, when tests over memory buffer analysis as well as polygon combine as 2 dissimilar geographical information systems services to apply on 2 mixed distributed operational management data sources. Sun & Li (2008) stated that outcomes demonstrate that the mobile grid computing technology based platform not simply generate steady outcomes as compared to making use of some traditional applications, however it also offers competent techniques for overall communication and operational management. This enhanced process management efficiency is influenced for the transfer of network processed data between data supplier, clients and servers (Sun & Li, 2008).

Gen (2010) stated that in case of distributed mobile information systems based grid model (DMGG) supported through mobile agent intended for real time downloading of spatial data and information. The arrangement of distributed grid computing supported through the mobile computing environment has high-quality load balance, enhanced processing competence as well as fewer loads on the overall network communication arrangement (Gen (b), 2010).

There are very small numbers of models present for enhancing the addition of Grid as well as mobile computing arrangement. Technology based system structure of these systems have employed the mobile agent technology methods for enhancing the methods of spatial data retrieval, access as well as processing (Tong & Shen, 2009.) & (Gen T. , 2010.). The incorporation of Mobile computing and Grid Computing Agents over communication and collaboration network has not been extremely analyzed intended for integrated system parameters similar to active workload judgments intended for every executing agent as well as dynamic system parameters judgments intended for every node of the grid computing arrangement in scenario of consumption idle memory, processor cycles and CPU time, etc. (Tong & Shen, 2009.) & (Singh & Bawa, 2012).

### 5.3- Challenges and Limitations in Mobile Grids

This section this research will present a deep and detailed analysis of some of main and fundamental challenges in application of Mobile Grid Computing architecture. Because of the wireless nature of the mobile devices as well as limitations of wireless collaboration and communications many unique issues and challenges need to recognize when developing a grid technology based application for such systems. Systems have to be proficient of regulating to frequent disconnections of some devices from the grid computing architecture. Huge distributed grid computing systems present new issues in task scheduling because of complicated task workload features as well as system features. Due to the many limitations that have to be recognized as well as the complex communications that are able to happen among dissimilar technology based resource allotment strategies; it meets numerous issues and aspects. A number of the needs as well as challenges of mobile grids computing arrangement are outlined below (Manvi, 2010):

### 5.3-1- Resource status monitoring

In case of technology based arrangement, monitoring involves the operations regarding gathering data and information associated to features as well as condition of available resources. Monitoring and assessment system have to be capable to offer information regarding the present state of different resources for example dynamic and static resources, as well as produce alarms when specific or significant events happen. It is responsible intended for assessing and monitoring the general performance and health of grid computing arrangement. To offer current technology based resources present status data, there is a device that is able to monitor either constantly or occasionally check the present status, when some specific event happens (Manvi, 2010). A number of tasks are performed to tackle this issue in a Grid computing arrangement. In this scenario Globus toolkit contains numerous monitoring tools and software those are known as the MDS or Monitoring and Discovery System tools (GLOBUS, 2012) & (Chung & Chang, 2009).

### 5.3.2- Resource status updating and communication:

To uphold current technology based resource's present status, a constant assessment and monitoring is required. The augment in amount of resource status deliverance of similar monitored explanation will use a great deal of bandwidth, formulating database size that is augmenting constantly with time. Therefore it is extremely much necessary to monitor simply related transformation and then communicate right away. There is a Grid Resource Information Monitoring toolkit or GRIM that is developed to support the continuously altering technology based resource states (Manvi, 2010).

### 5.3.3- Authentication and Authorization of device/user

As the Mobile Gird Computing system is developed through pervasive and wireless network, it will be harder to carry out authorization and authentications as well as to offer broad security methods. The verification of mobile grid devices as well as linked users as they connect to network offers extra issues and complexities in mobile based grids computing arrangements. To get entry to some mobile grid technology based services or resources, every system and device as well as client has to be conformed and authorized. Although numerous studies have assessed this issue of user identity confirmation in ad-hoc mobile grids computing arrangement, a small number of solutions have been approaching (Manvi, 2010). The present established identity administration method depends on the similar digital signature corticated in utilize on the traditional wired Internet based systems. This research has been performed by (Bhagyavati & Kurkovsky, 2004) and outlined humorous similar issues.

### 5.3.4- Resource description

It is the recognized procedure to explain the technology based resources to be shared in any network based arrangement. The heterogeneity of mobile grid computing structure devices, OSs they utilize, collaborative services they offer, data they signify requires a standard to explain the technology based resources so that some mobile grid computing arrangement is able to make use of the technology based services. A number of standards intended for technology based resource explanation (Litke, Skoutas, & Varvarigou, Mobile grid computing: Changes and Challenges of resource management in a mobile grid environment, 2004) comprise RDF, XML and OWL.

### 5.3.5- Resource discovery

The idea of technology based resource sharing in mobile grid computing arrangement is recognized through service detection methods those clearly plus seamlessly position accessible gird based services/resources all through grid based infrastructure upon client request. As client desired the mobile grid service as quicker as probable at practical cost, technology based service discovery turns out to be more significant aspect in grid based computing arrangement computing (Litke A. , Dimitrios, & Theodora, 2004). Therefore there is need for suitable indexing; publishing plus cataloguing of available shared resources turn out to be extremely essential. Technology based standards for example WSDL; GRDL etc. are obtainable intended for resource discovery in a much better way (Manvi, 2010).

### 5.3.6- Resource allocation

Through mobile technology based devices are more constrain in their memory, processing power as well as bandwidth, numerous clients struggle for these limited technology based resources. A number of financial marketplace models are favored to assign the resources on cost basis. In this scenario, appropriate network resource allocation job scheduling turns out to be extremely essential so as to offer well-organized service to grid computing users. A research based survey on technology based resource sharing in service oriented mobile grids is outlined by (Batista & Fonseca, 2009). Resource sharing has o be performed following the managerial strategies as well as the circumstances under those network devices will expand access to other systems (Manvi, 2010).

### 5.3.7- Routing of messages through the grid

This aspect is linked to the decrease of power consumption however by means of competence in the mobile grid routing of messages. A subsequent concern is the decrease of communication message latency (Manvi, 2010).

### 5.3.8- Power consumption

As mobile network devices are having numerous resource constrained, recognizing methods to minimize power utilization to enhance battery life as keeping satisfactory levels of mobile grid services service at better extent. This is mainly significant in wireless sensor grids computing arrangement, because of small size of the sensors as well as the issues regarding replacing those that fail. The research work outlined by (Ye, Heidemann, & Estrin, 2002) stated that regarding minimizing the power utilization as well as routing the data and information by means of the paths that minimize power reduction (Manvi, 2010).

**5.3.9- Mobility**

If the systems in the mobile grid are very much wireless, it will unfavorably influence on the time needed to carry out highly intensive tasks regarding computations. The mobility of the systems also reasons network instability issues those are leading to less effective quality of services. In case of a resource sharing atmosphere: for example wireless grid, mobility reasons loss of information resulting in possible invalid outputs. Manvi, (2010) stated that influence mobility in mobile grid computing arrangement plus outlined a structure that attempts to reduce the negative influences of mobility (Manvi, 2010).

**5.3.10- Information Security**

Security of the information mobile grid computing arrangement is also one of main. Because of wireless nature of the devices in the grid computing arrangement there are a great deal of issues and concerns regarding the safety and security of information over the gird arrangement (Manvi, 2010).

## Chapter 6: Conclusion and Recommendations

**6) Conclusion and Recommendations**

This research has offered a deep and detailed analysis of relationship between mobile and grid technology. This research has offered a comprehensive analysis of grid computing and mobile computing. For the sake I have outlined the main reasons for which we basically need to go for grid computing. Here this research has outlined the basic mapping between grid computing techniques on mobile domains. For this sake I have assessed how can we achieve grid computing services through application of mobile devices. This research has also assessed the existing grid computing models for mobile computing technology. This research has also offered a detailed analysis of its design, implementation, architecture and its limitations of mobile grid computing architecture. In this scenario, I have assessed the grid computing aspects regarding mobile's load balancing, high processing efficiency and less network communication.

Mobile grid technology is an emerging paradigm that is presently in its evolutionary stages. There is great deal of research and future working still pending in these areas. For example security in such a distributed arrangement is one of main aspect regarding designing mobile computing grid architecture. There is need for conformation and application of enhanced security management tools and methods.

Ultimately we expecting future, in which we will be linked through our homes or in the course of our wireless or mobile devices, are clients of grid technology. A number of have sketched the thoughts of performing games by means of grid platforms, or sales and purchase arrangements the idle power of our system to Grid Service suppliers, immediately as solar energy clients sell the idle power back to the energy grid. This enveloping idea of the grid is yet at least a decade away; however when it turns out to be a realism, telecommunications corporations will have the chance to be several of the most relevant players in it. The mobile grid technologies will fortune every area of life from household to business or production site to battle filed. The coming future is full of hope and support for offering a great deal of support and capability for better technology implementation.

## Chapter 6: References